\documentclass{elsart}
\usepackage{amssymb,amsfonts}
\usepackage{graphicx}
\newcommand{\trace}{\mathop{\rm Tr}\nolimits}

\newcommand{\supp}{\mathop{\rm supp}\nolimits}

\newcommand{\cH}{{\mathcal H}} 
\newcommand{\cT}{{\mathcal T}} 
\newcommand{\cR}{{\mathcal R}} 

\newcommand{\id}{\mathbb{I}}

\newcommand{\be}{\begin{equation}}
\newcommand{\ee}{\end{equation}}
\newcommand{\bea}{\begin{eqnarray}}
\newcommand{\eea}{\end{eqnarray}}
\newcommand{\beas}{\begin{eqnarray*}}
\newcommand{\eeas}{\end{eqnarray*}}
\newtheorem{definition}{Definition}
\newtheorem{theorem}{Theorem}
\newtheorem{lemma}{Lemma}

\newtheorem{proposition}{Proposition}

\newcount\minute
\newcount\hour
\def\currenttime{%
    \minute\time
    \hour\minute
    \divide\hour60
    \the\hour:\multiply\hour60\advance\minute-\hour\the\minute}
\begin{document}
\begin{frontmatter}
\title{Telescopic Relative Entropy--II\\ Triangle inequalities}
\author{Koenraad M.R.\ Audenaert}
\address{
Mathematics Department,
Royal Holloway, University of London, \\
Egham TW20 0EX, United Kingdom}
\ead{koenraad.audenaert@rhul.ac.uk}
\date{\today, \currenttime}
\begin{abstract}
In previous work, we have defined the telescopic relative entropy (TRE),
which is a regularisation of the quantum relative entropy
$S(\rho||\sigma)=\trace\rho(\log\rho-\log\sigma)$, by replacing the second argument $\sigma$ by a
convex combination of the first and the second argument, $\tau=a\rho+(1-a)\sigma$ and dividing
the result by $-\log a$. We also explored some basic properties of the TRE.
In this follow-up paper we state and prove two upper bounds on the variation of the TRE
when either the first or the second argument changes. These bounds are close in spirit to
a triangle inequality. For the ordinary relative entropy no such bounds are possible due to the fact
that the variation could be infinite.
\end{abstract}

\end{frontmatter}
\section{Introduction\label{sec:intro}}
The quantum relative entropy between two quantum states $\rho$ and $\sigma$,
$S(\rho||\sigma)=\trace\rho(\log\rho-\log\sigma)$, is a non-commutative generalisation
of the Kullback-Leibler distance between probability distributions and
is widely used as a distance measure
between quantum states \cite{ohya_petz}.
One of its main drawbacks, however, is that the relative entropy is infinite when
$\{\rho\} \not\ge \{\sigma\}$ (with $\{\rho\}$ denoting the projector on the support of $\rho$).
In particular, relative entropy is useless as a distance measure between pure states,
since it is infinite for pure $\rho$ and $\sigma$, unless
$\rho$ and $\sigma$ are exactly equal (in which case it always gives $0$).

To overcome this problem,
in \cite{TRE1} we introduced a regularisation of the relative entropy, which we call the
\textit{telescopic relative entropy} (TRE):
\begin{definition}
For fixed $a\in(0,1)$, the $a$-telescopic relative entropy between states $\rho$ and $\sigma$ is given by
\be
S_a(\rho||\sigma) := \frac{1}{-\log(a)} \,\,S(\rho||a\rho+(1-a)\sigma).
\ee
\end{definition}
We showed that the value of the TRE is always between $0$ and $1$;
$S_a(\rho||\sigma)=1$ if and only if $\rho\perp\sigma$.

Furthermore, we have defined the limits $a\to0$ and $a\to1$ and have shown that these limits exist and can be expressed in closed form:
\begin{theorem}\label{th:S01closed}
For any pair of states $\rho$, $\sigma$,
\bea
S_0(\rho||\sigma) &:=& \lim_{a\to 0}S_a(\rho||\sigma) =  1-\trace \rho\{\sigma\} \\
S_1(\rho||\sigma) &:=& \lim_{a\to 1}S_a(\rho||\sigma) = 1-\trace \sigma\{\rho\}.
\eea
\end{theorem}

The origin of the name `telescopic' is that the operation $\sigma\mapsto a\rho+(1-a)\sigma$ acts like a `telescope'
with `magnification factor' $1/(1-a)$, bringing the state $\sigma$ closer to the `vantage point' $\rho$
and bringing observed pairs of states $\sigma_i$ closer to each other.

The main result of the present paper (see Section \ref{sec:cont}) is the
establishing of two upper bounds on the variation of the TRE, when either one of the
arguments varies.
In some sense these bounds could be considered as close relatives of the triangle inequality,
while in another sense they could be considered as Fannes-type continuity inequalities.
The two inequalities have no counterpart
for the ordinary relative entropy, because the constants appearing in them
would have to be infinite. The existence of reasonable bounds for the TRE is due to the telescoping process.

In the next section we first collect the prerequisites, defining the basic notations and stating known relations for
the relative entropy and the right-derivative of the operator logarithm.
\section{Preliminaries\label{sec:pre}}
\subsection{Notations}
For any self-adjoint operator $X$ on a Hilbert space $\cH$,
we denote by $\supp X$ the support of $X$, i.e.\
the subspace of $\cH$ which is the orthogonal complement of $\ker X$, the kernel of $X$.
The projector on the support of $X$ will be denoted by $\{X\}$.

For any self-adjoint operator $X$, $X_+$ will denote the positive part $X_+ = (X+|X|)/2$.
It features in an expression
for the trace norm distance between states:
\be
T(\rho,\sigma) := \frac{1}{2}||\rho-\sigma||_1 = \trace(\rho-\sigma)_+.
\ee
The trace of the positive part has a variational characterisation as
$\trace X_+ = \max_P \trace XP$, where the maximisation is over all self-adjoint projectors.
Hence, for all such projectors $P$, $\trace XP\le \trace X_+$.

Two quantum states are mutually orthogonal, denoted $\rho\perp\sigma$, iff $\trace\rho\sigma=0$.
\subsection{Gradients}
We will need the gradients of the
relative entropy. Here we bring together all known facts.

The following integral representation of the logarithm lies at the basis of much of the subsequent
treatment. For $x>0$, we have
\be
\log x = \int_0^\infty ds \left(\frac{1}{1+s}-\frac{1}{x+s}\right).\label{eq:intlog}
\ee
This immediately provides an integral representation for the ordinary relative entropy:
\bea
S(\rho||\sigma) &=& -\int_0^\infty ds\,
\trace \rho [(\rho+s)^{-1}-(\sigma+s)^{-1}] \label{eq:int1} \\
&=& -\int_0^\infty ds\,
\trace \rho (\rho+s)^{-1}\,\, (\sigma-\rho)\,\, (\sigma+s)^{-1}. \label{eq:int2}
\eea
Likewise, we get similar expressions for the telescopic relative entropy:
\bea
\lefteqn{S_a(\rho||\sigma)} \nonumber \\
&=& \frac{1}{\log a}\,\,\int_0^\infty ds\,
\trace \rho [(\rho+s)^{-1}-(a\rho+(1-a)\sigma+s)^{-1}] \label{eq:int1a} \\
&=& \frac{1}{\log a}\,\,\int_0^\infty ds\,
\trace \rho (\rho+s)^{-1}\,\, (1-a)(\sigma-\rho)\,\, (a\rho+(1-a)\sigma+s)^{-1}. \label{eq:int2a}
\eea

Another integral we will encounter is $\int_0^\infty ds \,\,\,x/(x+s)^2$.
For $x=0$, the integral obviously gives $0$. For $x>0$ it gives $1$.
Hence
\be
\int_0^\infty ds\, (\rho+s)^{-1}\,\rho \, (\rho+s)^{-1} = \{\rho\}.\label{eq:intproj}
\ee

The gradient of the relative entropy w.r.t.\ its first argument is defined through the relation
\beas
\trace \Delta \,\,\nabla_1 S(A||B)
&=& \frac{d}{dt}\Bigg|_{t=0} S(A+t\Delta||B) \\
&=& \trace \Delta (\log(A)-\log(B)) + \trace A \frac{d}{dt}\Bigg|_{t=0} \log(A+t\Delta).
\eeas
The gradient of the relative entropy w.r.t.\ its second argument is defined similarly through
\beas
\trace \Delta \,\,\nabla_2 S(A||B)
&=& \frac{d}{dt}\Bigg|_{t=0} S(A||B+t\Delta) \\
&=& -\trace A \frac{d}{dt}\Bigg|_{t=0} \log(B+t\Delta).
\eeas

Hence, to find explicit expressions, the derivative of the logarithm is needed.
From integral representation (\ref{eq:intlog})
we get
$$
\frac{d}{dt}\Bigg|_{t=0} \log(A+t\Delta)
=\int_0^\infty ds\,\,(A+s\id)^{-1} \Delta (A+s\id)^{-1}.
$$
It will be useful to introduce the following linear map, for $A\ge0$:
\be
\cT_A(\Delta) = \int_0^\infty ds\,\,(A+s\id)^{-1} \Delta (A+s\id)^{-1}.
\ee
Thus
\be
\frac{d}{dt}\Bigg|_{t=0} \log(A+t\Delta) = \cT_A(\Delta).
\ee
It's easy to check that for $A\ge0$, $\cT_A(A) = \{A\}$.
Thus, for $A>0$, we have $\cT_A(A)=\id$.

From the integral representation it also follows that, for any self-adjoint $A$, $\cT_A$ preserves the
positive semidefinite order: if $X\le Y$, then $\cT_A(X)\le\cT_A(Y)$.
By cyclicity of the trace, we see that the map $\cT_A$ is self-adjoint:
$\trace B\cT_A(\Delta) = \trace\Delta\cT_A(B)$.
Moreover, the map is positive semi-definite, in the sense that
$\trace\Delta\cT_A(\Delta)$ is positive for any self-adjoint $\Delta$.
This follows from the integral representation
and the fact that for positive $X$ and self-adjoint $Y$, $\trace XYXY = \trace(X^{1/2}YX^{1/2})^2\ge0$.

Further properties are discussed in \cite{lieb73}.
In particular, the map $(A,X)\mapsto \trace X^*\cT_A(X)$, for $A\ge0$ and any $X$,
is jointly convex in $A$ and $X$ (\cite{lieb73}, Theorem 3).

Using these properties of $\cT_A$ one easily obtains (see also \cite{ohya_petz}, Chapter 3):
\begin{lemma}
Let $A$ and $B$ be positive operators.
The gradient of $S(A||B)=\trace A(\log A-\log B)$ w.r.t.\ $A$ is given by
\be
\nabla_1 S(A||B) = \log A-\log B+\{A\}.
\ee
The gradient w.r.t.\ $B$ is given by
\be
\nabla_2 S(A||B) = -\cT_B(A).
\ee
\end{lemma}
The corresponding statement for the telescopic relative entropy is
\begin{lemma}
Let $A$ and $B$ be positive semidefinite operators.
The gradient of $S_a(A||B)$ w.r.t.\ $A$ is given by
\be
\nabla_1 S_a(A||B) = \frac{1}{-\log a}(\log A-\log C+\{A\}-a\cT_C(A)),
\ee
with $C=aA+(1-a)B$.
The gradient w.r.t.\ $B$ is given by
\be
\nabla_2 S_a(A||B) = -\frac{1-a}{-\log a}\cT_B(A).
\ee
\end{lemma}

\bigskip

Having defined the linear operator $\cT$ via the first derivative of the logarithm, we can also define
a quadratic operator $\cR$ via the second derivative \cite{lieb73}. For $A\ge0$ and $\Delta$ self-adjoint,
\be
\cR_A(\Delta) = -\frac{d^2}{dt^2}\Bigg|_{t=0}\log(A+t\Delta).
\ee
A simple calculation using the integral representation of the first derivative yields the
integral representation
\be
\cR_A(\Delta) = 2\int_0^\infty ds\,\,(A+s\id)^{-1}\Delta(A+s\id)^{-1}\Delta(A+s\id)^{-1}.
\ee

\subsection{Basic properties of Telescopic Relative Entropy\label{sec:basic}}
We have shown in \cite{TRE1} that the value of the telescopic relative entropy
is always between $0$ and $1$, even for non-faithful states.
Furthermore, it inherits many desirable properties from the ordinary relative entropy,
like positivity, the fact that it is only zero when $\rho$ and $\sigma$ are equal (provided $a>0$),
joint convexity in its arguments, and monotonicity under CPT maps.

The following identities are straightforward:
\bea
b S_a(X||Y) &=& S_a(bX||bY)\\
S_a(bX||cX) &=& S_a(b||c).
\eea
As we do not restrict the arguments of the telescopic relative entropy to states,
the definition of TRE is also applicable to non-negative scalars:
\be
S_a(b||c) = \frac{b(\log b - \log(ab+(1-a)c))}{-\log a}.
\ee
In particular, we have
\be
S_a(b||0)=b,\qquad S_a(0||c)=0.
\ee

Both the ordinary relative entropy and the TRE satisfy certain monotonicity properties when applied to
non-normalised positive operators:
\begin{lemma}\label{lem:monoboth}
For $A,B,X\ge0$,
\bea
S(A+X||B+X)&\le& S(A||B)\\
S_a(A+X||B+X)&\le& S_a(A||B)\\
S(A||B+X)&\le& S(A||B)\label{eq:monoboth}\\
S_a(A||B+X)&\le& S_a(A||B).
\eea
\end{lemma}
\textbf{Proof.}
The former two inequalities follow from joint convexity of $S$ and $S_a$:
\beas
S_a(A+X||B+X) &=& 2S_a((A+X)/2||(B+X)/2) \\
&\le& S_a(A||B)+S_a(X||X) \\
&=& S_a(A||B).
\eeas
The latter two inequalities follow from operator monotonicity of the logarithm.
\qed

\section{Main results\label{sec:cont}}
In this section,
we present a number of highly non-trivial inequalities concerning the telescopic relative entropy.
We like to point out here that these inequalities are just as much about the ordinary relative entropy, applied
in the setting where $\sigma$ is a convex combination of $\rho$ and another state, and are therefore of interest
regardless whether one wishes to use the telescopic relative entropy as an independent concept or not.

While the telescopic relative entropy shares many properties with the ordinary relative entropy, and improves
on certain undesired properties, just like the relative entropy
it does not satisfy a triangle inequality in the strictest sense:
$S_a(\rho||\tau)\not\le S_a(\rho||\sigma)+S_a(\sigma||\tau)$.
However, due to the telescoping,
inequalities can be proven that at least come close in spirit to a triangle inequality.

Here we prove bounds on the difference between two telescopic relative entropies,
the first between the distances from $\rho$ to $\tau_1$ and to $\tau_2$, respectively,
in terms of the trace norm distance between
$\tau_1$ and $\tau_2$; the second between the distances from $\rho_1$ to $\tau$ and $\rho_2$ to $\tau$, in terms
of the trace norm distance between $\rho_1$ and $\rho_2$. These two bounds prove the continuity of the
telescopic relative entropy in both of its arguments, in the sense of Fannes.

First we state a triangle inequality w.r.t.\ the first argument:
\begin{theorem}\label{th:triangle1}
For $a\in(0,1)$, and for states $\rho_1,\rho_2,\sigma$ such that $T(\rho_1,\rho_2)=t$,
\bea
|S_a(\rho_1||\sigma)-S_a(\rho_2||\sigma)|
&\le& 1-S_a(1-t||0)-S_a(t||1) \nonumber \\
&=& t-S_a(t||1).
\eea
\end{theorem}
It is easily verified that equality is achieved for $\rho_1\perp\sigma$ and $\rho_2=t\sigma+(1-t)\rho_1$.
For the ordinary relative entropy no such bound is possible, as can be seen by taking two different
pure states for $\rho$ and $\sigma_1$, and a mixed state for $\sigma_2$: for such a choice
the difference $|S(\rho||\sigma_1)-S(\rho||\sigma_2)|$ becomes infinite.

The proof of this theorem relies on the following proposition, which may be of independent interest:
\begin{proposition}\label{prop:rbts2}
For $A,B,X$ positive operators, with $b=\trace B$ and $x=\trace X$,
\be
S(A||A+X)\ge S(A+B||A+B+X) \ge S(A||A+X)+S(b||b+x)
\ee
\end{proposition}

Next, we state a triangle inequality w.r.t.\ the second argument:
\begin{theorem}\label{th:triangle2}
For $a\in(0,1)$, and for states $\rho,\sigma_1,\sigma_2$ such that $T(\sigma_1,\sigma_2)=t$,
\bea
|S_a(\rho||\sigma_1) - S_a(\rho||\sigma_2)|
&\le& 1-S_a(1||t) \\
&\le& \frac{1-a}{-a\log a}\,t.
\eea
\end{theorem}
When $\rho\perp\sigma_1$ and $\sigma_2=t\rho+(1-t)\sigma_1$, equality is achieved.
This shows that the inequality is sharp, for any $a$ and $t$.

Again, for the ordinary relative entropy no such bound is possible, as can be seen by taking two different
pure states for $\rho$ and $\sigma_1$, and a mixed state for $\sigma_2$: for such a choice
the difference $|S(\rho||\sigma_1)-S(\rho||\sigma_2)|$ becomes infinite.

Note that the coefficient $(1-a)/(-a\log a)$ is always greater than or equal to $1$.
It tends to $+\infty$ in the limit $a\to 0$ and to $1$ in the limit $a\to 1$.
This implies in particular that $|S_1(\rho||\sigma_1)-S_1(\rho||\sigma_2)| \le T(\sigma_1,\sigma_2)$.
That can also be seen to follow from Theorem \ref{th:S01closed}, as
\beas
|S_1(\rho||\sigma_1)-S_1(\rho||\sigma_2)|
&=& |\trace\{\rho\}(\sigma_1-\sigma_2)| \\
&\le& \trace(\sigma_1-\sigma_2)_+ \\
&=&T(\sigma_1,\sigma_2).
\eeas

The proof of Theorem \ref{th:triangle2} relies on the following proposition,
which is a counterpart of Proposition \ref{prop:rbts2}:
\begin{proposition}\label{prop:rbts}
For $A,B,X\ge0$, with $b=\trace B$ and $x=\trace X$,
\be
S(X||A+X)\ge S(X||A+B+X) \ge S(X||A+X) + S(x||b+x).
\ee
Equivalently, for every state $\rho$ and all $A,B\ge0$, with $b=\trace B$,
\be
0\le \trace\rho(\log(\rho+A+B)-\log(\rho+A)) \le \log(1+b).\label{eq:rbts}
\ee
\end{proposition}

Finally, we consider the remaining case, $S_0$ and $S_1$:
\begin{theorem}
The telescopic relative entropy $S_0(\rho||\sigma)$ is continuous in $\rho$ but discontinuous in $\sigma$.
For $S_1(\rho||\sigma)$ the opposite situation holds.
\end{theorem}
\textbf{Proof.}
This can be seen immediately from the closed form expressions of Theorem
\ref{th:S01closed}. Linearity in one argument obviously implies continuity in that argument.
On the other hand, the function that maps a state to the projector on its support is
discontinuous, and this shows discontinuity in the other argument.
\qed
\section{Proofs}
In this section we will prove the stated triangle inequalities.
The centerpiece in these proofs is the following proposition:
\begin{proposition}\label{prop:Tderiv}
For $A,B,X\ge0$, with $b=\trace B$, $x=\trace X$,
\be
0 \le \trace X \cT_{A+X}(X) - \trace X \cT_{A+B+X} (X) \le \frac{bx}{b+x}.\label{eq:Tderiv}
\ee
\end{proposition}

To prove this we first need some lemmas.
\begin{lemma}\label{lem:lieb1}
For $A,B\ge0$,
\be
\trace(A+B)\cR_{A+B}(A)\ge \trace A\cT_{A+B}(A).
\ee
\end{lemma}
\textbf{Proof.}
It has been proven in \cite{lieb73} that
for $A,B\ge0$ and $K,M$ self-adjoint,
$$
-\trace B\cR_A(K)+2\trace M\cT_A(K) \le \trace M\cT_B(M).
$$
The inequality of the lemma follows by replacing $A$ and $B$ by $A+B$, and $K$ and $M$ by $A$.
\qed

\begin{lemma}\label{lem:lieb2}
For $A,B\ge0$,
\be
\cR_{A+B}(A)\le\id.
\ee
\end{lemma}
\textbf{Proof.}
We use the integral representations of $\cT$ and $\cR$.
Since $A+B+s\id\ge B$, we have $(A+B+s\id)^{-1}\le B^{-1}$ and $B(A+B+s\id)^{-1}B\le B$.
Therefore,
\beas
\cR_{A+B}(B) &=& 2\int_0^\infty ds\,\,(A+B+s\id)^{-1}\,\,B(A+B+s\id)^{-1}B\,\,(A+B+s\id)^{-1} \\
&\le&2\int_0^\infty ds\,\,(A+B+s\id)^{-1}\,\,B\,\,(A+B+s\id)^{-1} \\
&=& 2\cT_{A+B}(B).
\eeas
It has been proven in \cite{lieb73} that, for $A\ge0$ and $\Delta$ self-adjoint,
$$
\id + 2\cT_A(\Delta)+\cR_A(\Delta) =\cR_A(A+\Delta)\ge0.
$$
Therefore,
\beas
\cR_{A+B}(A+B-B)&=&\id+2\cT_{A+B}(-B)+\cR_{A+B}(-B) \\
&=& \id-2\cT_{A+B}(B)+\cR_{A+B}(B).
\eeas
Thus, indeed, $\cR_{A+B}(A)\le\id$.
\qed

\begin{lemma}\label{lem:f}
Let $f(t)$ be a real-valued convex function on $[0,1]$.
If, moreover, $f(0)\le0$ and $f(0)\le f'(0)$,
then $\forall t\in[0,1], f(0)\le (1-t)f(t)$.
\end{lemma}
\textbf{Proof.}
Since $f(0)\le0$, for all $t\in[0,1]$ we have $f(0)/(1-t) \le f(0) \le f'(0)$.
Multiplying both sides by $t(1-t)$ yields $tf(0)\le t(1-t)f'(0)$.
Adding $(1-t)f(0)$ to both sides gives
$f(0) \le t(1-t)f'(0) +(1-t)f(0) = (1-t)(f(0)+tf'(0))$.
By convexity of $f$, $f(0)+tf'(0)$ is a lower bound on $f(t)$,
and the inequality of the lemma follows.
\qed

\textbf{Proof of Proposition \ref{prop:Tderiv}.}
The first inequality in (\ref{eq:Tderiv}) easily follows from the fact that $x\mapsto 1/x$ is
operator monotone decreasing together with the identity
$$
\trace X \cT_A(X) = \int_0^\infty d\lambda \,\,\trace(X^{1/2}(A+\lambda\id)^{-1} X^{1/2})^2,
$$
and monotonicity of $\trace X^2$.
The second inequality involves more work.

Let us introduce an operator $G$ such that $G\ge\rho$,
and consider the function
$$
f(t)=\trace\rho\cT_{t\sigma + (1-t)G}(\rho)-1.
$$
We will first show that $(1-t)f(t)\ge f(0)$ for $0\le t\le 1$.

The derivative $f'(0)$ can be calculated explicitly from the integral representation of $\cT$:
\beas
f'(0) &=& \frac{d}{dt}\Bigg|_{t=0} \trace\rho\cT_{t\sigma+(1-t)G}(\rho) \\
&=& \frac{d}{dt}\Bigg|_{t=0} \int_0^\infty dx\,\,
\trace\rho (G+t(\sigma-G)+x\id)^{-1}\rho(G+t(\sigma-G)+x\id)^{-1} \\
&=& -\int_0^\infty dx\,\,\trace[\rho (G+x\id)^{-1}(\sigma-G)(G+x\id)^{-1}\rho(G+x\id)^{-1} \\
&&\qquad\qquad +\rho (G+x\id)^{-1}\rho(G+x\id)^{-1}(\sigma-G)(G+x\id)^{-1}] \\
&=& \trace(G-\sigma)\cR_G(\rho).
\eeas

By combining the inequalities of Lemma's \ref{lem:lieb1} and \ref{lem:lieb2} we obtain
$$
\trace(G-\sigma)\cR_G(\rho) \ge \trace \rho\cT_G(\rho)-1,
$$
which proves that $f'(0)\ge f(0)$.
In $t=0$, $f$ takes the value $\trace\rho\cT_G(\rho)-1$, which is non-positive, since
$\cT_G(\rho) \le \cT_G(G)=\id$. Thus $f(0)\le0$.
Moreover, by convexity of the map $(\rho,G)\mapsto \trace\rho\cT_G(\rho)$, $f(t)$ is convex.
By Lemma \ref{lem:f} these three statements imply that $(1-t)f(t)\ge f(0)$, for $0\le t\le 1$, i.e.\ the minimum
of $(1-t)f(t)$ over $[0,1]$ occurs in $t=0$.

Now let $t=b/(b+x)$, which indeed takes values in the interval
$[0,1]$, and is $0$ iff $b=0$. Then $1-t=x/(b+x)$. Also, let $G=\rho+A/x$.
With these substitutions, we get
$$
x(1-t)f(t) = \trace x\rho\cT_{b\sigma + x\rho+A}(x\rho)-\frac{x^2}{b+x},
$$
and we therefore find that this expression is minimal for $b=0$.
That is,
$$
\trace x\rho\cT_{b\sigma + x\rho+A}(x\rho)-\frac{x^2}{b+x} \ge
\trace x\rho\cT_{x\rho+A}(x\rho)-x,
$$
or, after rearranging terms,
$$
\trace x\rho\cT_{A+x\rho}(x\rho) - \trace x\rho\cT_{A+b\sigma+x\rho}(x\rho) \le \frac{bx}{b+x}.
$$
By substituting $X=x\rho$ and $B=b\sigma$ we obtain the second inequality of the proposition.
\qed

\bigskip

This proposition now allows us to prove Propositions \ref{prop:rbts2} and \ref{prop:rbts}, and subsequently
Theorems \ref{th:triangle1} and \ref{th:triangle2}.

\textbf{Proof of Proposition \ref{prop:rbts2}.}
We want to show
$$
S(A||A+X)\ge S(A+B||A+B+X) \ge S(A||A+X)+S(b||b+x).
$$
The first inequality is Lemma \ref{lem:monoboth}.

The second inequality is proven using the second inequality in (\ref{eq:Tderiv}):
$$
\trace X \cT_{A+X}(X) - \trace X \cT_{A+B+X} (X) \le \frac{bx}{b+x}.
$$
On replacing $X$ by $tX$ and dividing both sides by $t$, we get
\be
\trace X \cT_{A+B+tX} (tX) - \trace X \cT_{A+tX}(tX) \ge -\frac{bx}{b+tx}.\label{eq:temp}
\ee
When $A>0$, $A+B+tX$ and $A+tX$ are positive for all $t$, so that
$$
\cT_{A+B+tX} (A+B+tX) = \cT_{A+tX} (A+tX) = \id.
$$
Therefore,
\beas
\lefteqn{\trace X \cT_{A+B+tX} (tX) - \trace X \cT_{A+tX}(tX)} \\
&=& -\trace X \cT_{A+B+tX} (A+B) + \trace X \cT_{A+tX}(A) \\
&=& -\trace (A+B) \cT_{A+B+tX} (X) + \trace A \cT_{A+tX}(X) \\
&=& \frac{d}{dt}\left(-\trace(A+B)\log(A+B+tX) + \trace A\log(A+tX)\right).
\eeas
Thus the inequality (\ref{eq:temp}) becomes
$$
\frac{d}{dt}\left(-\trace(A+B)\log(A+B+tX) + \trace A\log(A+tX)\right)
\ge -\frac{bx}{b+tx},
$$
for all $t\ge 0$.
Integrating over $t\in[0,1]$ this turns into
\beas
\lefteqn{(-\trace(A+B)\log(A+B+X)+\trace(A+B)\log(A+B))} \\
&&\mbox{ } + (\trace A\log(A+X)-\trace A\log A) \\
&\ge& \int_0^1 dt\,\,\frac{-bx}{b+tx} = b(\log b-\log(b+x)),
\eeas
which is the inequality of the proposition.

By a standard continuity argument, this is also true for $A\ge0$.
\qed

\textbf{Proof of Theorem \ref{th:triangle1}.}\\
W.l.o.g.\ we assume that $S_a(\rho_1||\sigma)$ is not less than $S_a(\rho_2||\sigma)$, so that the absolute
value signs can be removed. Changing signs, we will prove
$$
S_a(\rho_2||\sigma) - S_a(\rho_1||\sigma) \ge S_a(t||1)-t.
$$

Introduce $\Delta=\rho_2-\rho_1$, then $t= T(\rho_1,\rho_2)=\trace\Delta_+$.
Noting that $\rho_2=\rho_1+\Delta=\rho_1+\Delta_+ - \Delta_-$, Proposition \ref{prop:rbts2} yields
\beas
\lefteqn{S(\rho_2||\rho_2+\frac{1-a}{a}\sigma) - S(\rho_1||\rho_1+\frac{1-a}{a}\sigma)} \\
&\ge& S(\rho_1+\Delta_+||\rho_1+\Delta_+ +\frac{1-a}{a}\sigma) - S(\rho_1||\rho_1+\frac{1-a}{a}\sigma) \\
&\ge& S(t||t+(1-a)/a).
\eeas
Here we successively used the first inequality of Proposition \ref{prop:rbts2} with $B=\Delta_-$,
and its second one with $B=\Delta_+$.
After some elementary algebra concerning the constant $a$, we get the required inequality.
\qed

\textbf{Proof of Proposition \ref{prop:rbts}.}
We proceed in similar fashion as in the proof of Proposition \ref{prop:rbts2}.
Using the identity $\cT_X(X)=\id$ for $X>0$, the inequality of Proposition \ref{prop:Tderiv} is equivalent with
$$
\trace X(\cT_{A+B+X}(A+B) - \cT_{A+X}(A)) \le \frac{bx}{b+x}.
$$
Replacing $A$ by $tA$ and $B$ by $tB$ and dividing both sides by $t$ gives
$$
\trace X(\cT_{t(A+B)+X}(A+B) - \cT_{tA+X}(A)) \le \frac{bx}{tb+x}.
$$
Integrating w.r.t.\ $t$ over $[0,1]$ then yields the inequality of Proposition \ref{prop:rbts}.
\qed

\textbf{Proof of Theorem \ref{th:triangle2}.}
With $s=(1-a)/a$, and $\tau_i = a\rho+(1-a)\sigma_i$,
\beas
S(\rho||\tau_1)-S(\rho||\tau_2)
&=& \trace\rho(\log(a\rho+(1-a)\sigma_2)-\log(a\rho+(1-a)\sigma_1)) \\
&=& \trace\rho(\log(\rho+s\sigma_2)-\log(\rho+s\sigma_1)) \\
&=& \trace\rho(\log(\rho+s\sigma_1+s(\sigma_2-\sigma_1))-\log(\rho+s\sigma_1)) \\
&\le& \trace\rho(\log(\rho+s\sigma_1+s(\sigma_2-\sigma_1)_+)-\log(\rho+s\sigma_1)).
\eeas
In the last line we used the operator monotonicity of the logarithm and the fact that $X_+\ge X$.
With the identification $A=s\sigma_1$ and $B=s(\sigma_2-\sigma_1)_+$, inequality (\ref{eq:rbts}) yields
\beas
S_a(\rho||\sigma_1)-S_a(\rho||\sigma_2)
&\le& \trace\rho(\log(\rho+A+B)-\log(\rho+A))/(-\log a) \\
&\le& \log(1+s\trace(\sigma_2-\sigma_1)_+)/(-\log a) \\
&=& (\log(a+(1-a)T(\sigma_1,\sigma_2))-\log(a))/(-\log a).
\eeas
\qed

\begin{ack}
A substantial part of this work was done at the Institut Mittag-Leffler, Djurs\-holm (Sweden),
during an extended stay at its Fall 2010 Semester on Quantum Information Theory.
I also acknowledge conversations with R.\ Werner, J.\ Oppenheim, B.\ Nachtergaele,
F.\ Verstraete, M-B.\ Ruskai, M.\ Shirokov and the rest of the
IML crowd at the V\"ardshus.
\end{ack}

\end{document}